# Efficacy of Hydroxychloroquine as Prophylaxis for Covid-19

Márcio Watanabe
Universidade Federal Fluminense

## Abstract

Limitations in the design of the experiment of Boulware et al[1] are considered in Cohen[2]. They are not subject to correction but they are reported for readers' consideration. However, they made an analysis for the incidence based on Fisher's hypothesis test for means while they published detailed time dependent data which were not analyzed, disregarding an important information. Here we make the analyses with this time dependent data adopting a simple regression analysis.

We conclude their randomized, double-blind, placebo-controlled trial presents statistical evidence, at 99% confidence level, that the treatment of Covid-19 patients with hydroxychloroquine is effective in reducing the appearance of symptoms if used before or right after exposure to the virus. For 0 to 2 days after exposure to virus, the estimated relative reduction in symptomatic outcomes is 72% after 0 days, 48.9% after 1 day and 29.3% after 2 days. For 3 days after exposure, the estimated relative reduction is 15.7% but results are not statistically conclusive and for 4 or more days after exposure there is no statistical evidence that hydroxychloroquine is effective in reducing the appearance of symptoms.

Our results show that the time elapsed between infection and the beginning of treatment is crucial for the efficacy of hydroxychloroquine as a treatment to Covid-19.

## 1. Introduction

The novel coronavirus disease, Covid-19, caused by the virus SARS-coV-2 has caused a major pandemic worldwide. No vaccines or specific treatment were available until June 2020. Many existing medicines have been tested to treat patients, mostly in hospitalized patients under more severe clinical conditions. Hydroxychloroquine is the most discussed of these drugs and several studies have pointed to different outcomes.[3] Although some randomized clinical trials have shown the inefficacy of hydroxychloroquine treatment to prevent death in hospitalized patients,[4] the benefits to less severe patients in the beginning of the disease and as a pre-exposure or post-exposure prophylaxis is still under discussion. Boulware et al.[1] was the first randomized clinical trial which has tested hydroxychloroquine as a prophylaxis treatment to Covid-19. The authors of the study concluded there is no statistical evidence of hydroxychloroquine's efficacy when compared to placebo results. Besides some limitations in their experimental design, as pointed by Cohen,[2] up to 23th of June 2020, when our study was finished, their results were the most reliable information available about hydroxychloroquine treatment for Covid-19 as a prophylaxis. However, correctable inaccuracies in their estimates remained without revision.

In the section Methods, we discuss some of these inaccuracies and present the corrections. Furthermore, in the section Results, we provide an original analysis of a more detailed time series data, available at their supplementary appendix.[1]

## 2 – Methods

The randomized trial of Boulware et al.[1] aims to test the treatment of Covid-19 infected patients for 5 days with hydroxychloroquine as a prophylaxis, measuring its effect by a possible reduction on the incidence of symptomatic outcomes when compared to results from a placebo group. Adult patients who had epidemiological linkage with Covid-19 confirmed patients were included if exposure was within 4 days at the beginning of the experiment. Initially 921 asymptomatic participants were randomly assigned to treatment or placebo groups but 100 presented symptoms at the day of beginning of experiment (day 1) and were excluded. The primary measure of effect was incidence of Covid-19 compatible symptoms with laboratory or clinical confirmation within 14 days. In the placebo group, 58 of 407 participants (14.3%) presented symptoms from day 2 to day 14, while 49 of 414 participants (11.8%) of the treatment group presented symptoms in the same period. No serious adverse reactions were reported. A two-tailed Fisher exact test was used to obtain a p-value of 0.351 and at 95% confidence level, they have concluded hydroxychloroquine did not prevent illness compatible with Covid-19 when used as a postexposure prophylaxis. Next, we make some considerations on their statistical analysis to corroborate with the different choices we make at section Results. Limitations on the design of their experiment are discussed by Cohen.[2]

### 2.1 - Two-tailed test

The drug in test is known to have antiviral and anti-inflammatory activities and to effectively inhibit SARS-coV-2 in vitro.[5]

When applying a hypothesis test, the statistician responsible for the data analysis must define whether to use a two-tailed test or a one-tailed test. This definition typically depends on the alternative hypothesis to be tested. In the present problem, the question to be answered is whether the treatment is able to reduce the incidence of symptomatic patients in contrast to the null hypothesis H0 that it will be as good as placebo. Thus, a one-tailed test is the natural choice to the alternative hypothesis H1 in this problem, in contrast to the two-tailed test adopted in their study.[1]

We emphasize this choice has serious implications in both interpretation and quantitative results. In the adopted two-tailed test, if the test result leads to H0 rejection, the conclusion would be that the treatment with hydroxychloroquine presents a different result than the placebo, where the true incidence could be either higher or lower. In this two-tailed case, it would not be possible to conclude directly that the rejection of H0 implies the efficacy of the treatment, despite the observed incidences favoring the hydroxychloroquine group. Another important aspect to consider is that the use of the two-tailed test favors null hypothesis to be not rejected in comparison with the one-tailed test. As an example, the p-value obtained in the article by a two-tailed test is 35.1% while the p-value obtained by the same method but with a one-tailed test is 17.8% (for complete data see Boulware et al[1] ).

**2.2 - Treatment definition**

Precise treatment group definition is necessary to a proper conduction of the statistical analysis in a clinical trial. In Boulware et al., the treatment group has been apparently defined as the group of randomly selected patients which received hydroxychloroquine for five consecutive days in the following scheme: 800mg once, 600mg 6 to 8 hours later, then 600mg daily for 4 days.[1]

Treatment effect has been calculated from the incidence differences of symptomatic patients between day 14 and day 1. Note that of 921 eligible patients on day 0, 100 became symptomatic on day 1 and were excluded. However, patients who became symptomatic on days 2, 3, 4 and 5 were included in the calculation of the treatment effect. Note this subgroup of patients, although included in the analysis by the authors, had not completed the entire treatment when the response variable (symptomatic or asymptomatic) had been defined and once a patient had symptoms it is no longer possible to change his status.

Note that a large proportion of the symptomatic patients in the study have presented first symptoms between day 2 and day 5, the period of treatment application (see Table 1 below). These patients did not have the full effect of the treatment. Using a similar logic that had made the authors exclude symptomatic patients in day 1, we should exclude patients with symptoms in days 2, 3, 4 and 5, otherwise treatment definition and statistical method should be changed.

Therefore, in order to have a homogeneous sample of patients taking the same dose, an adequate analysis should measure the differences in incidences from days 14 to 5, and not from 14 to day 1. The high sensitivity of the response variable chosen by the researchers (symptomatic or asymptomatic) is another reason why people who had symptoms before finishing treatment should be excluded from measuring the effects on the test.

The sample obtained considering only patients who presented first symptoms after the fifth day is far from ideal because there can be many patients for whom a long time has passed from the day of infection to the day of the end of treatment. Unfortunately, the best way to avoid this problem should have been adopted in the design of the experiment. An alternative solution to include patients who have not used full treatment before the onset of symptoms is to use a statistical method more appropriate to a heterogeneous sample such as a regression method (see section 3).

Hence, to test the effect of a complete 5 days treatment we should compare only the incidence differences between the groups from day 14 to day 5 (patients who have presented symptoms only after the end of complete prophylaxis). Raw data are not available, but with figure 2 of their article we can obtain the following approximation[1]:

Table 1: Percentage of symptomatic patients

|  | Day 5 | Day 14 | Difference | Symptomatic | Asymptomatic |
|---|---|---|---|---|---|
| Treatment | ≈ 7.7% | 11.8% | 4.1% | a=17=49-32 | b=365 |
| Placebo | ≈ 7.4% | 14.3% | 6.9% | c=28=58-30 | d=349 |

Applying Fisher's exact test to this approximate data, for a one-tailed test, we obtain a p-value of 5.6%, which is considerably different from the 35.1% of their article.

## 2.3 - Measure of efficacy

Their study uses absolute difference of the proportions of symptomatic patients between treatment and placebo groups as its response variable. However, this variable is not explanatory because it is not robust to inclusion of asymptomatic uninfected patients. First, observe that there are many more asymptomatic than symptomatic patients in the sample (714 and 107). Second, most patients in the study were not tested, so a significant proportion of this sample should not be infected, decreasing a possible reduction in the absolute incidence of treatment effect and artificially increasing the sample size and, as a consequence, the reliability of the statistics in the study.

The real number of asymptomatic infected patients is unknown in each group, and Fisher's exact test cannot be applied to this data, because the test must be applied to symptomatic infected versus asymptomatic infected in each group, since asymptomatic uninfected are not sensitive to any treatment, no matter the treatment is effective or not.

Hence, in order to verify whether the difference between rates of symptomatic patients in the two groups is significant, the ideal sample should have included only confirmed infected patients. This major problem in the design of the experiment can not be changed, but it can be significantly minimized by selecting a statistic to measure treatment effect which is less affected by the inclusion of uninfected asymptomatic patients in the sample. The relative difference (Rd) is a more appropriate measure in this case:

$$Rd = \frac{\text{rate of symptomatic at hydroxychloroquine group} - \text{rate of symptomatic at placebo group}}{\text{rate of symptomatic at placebo group}}$$

The relative difference can be interpreted as the negative of a measure of the percentage of treatment effectiveness. That is, define Treatment efficacy = -Rd. Suppose for example the relative difference Rd = -0.30. Then, hydroxychloroquine treatment efficacy is 30% in this case, which implies, on average, 30 out of 100 patients who would have presented symptoms if they have taken placebo will no longer develop symptoms if they take hydroxychloroquine. This variable is more informative and it is far less affected by the unknown proportion of uninfected patients in the sample.

To illustrate this, consider the following example: Let Sh and Sp be the number of symptomatic patients in treatment and placebo groups respectively. Suppose N is the total number of patients in each group. Then in this case, absolute difference of incidences is given by |Sh-Sp|/N and relative difference Rd={(Sh/N)-(Sp/N)}/{Sp/N}= (Sh-Sp)/Sp. If we add M=N uninfected patients to each group, then the absolute difference will be half of the original value interfering in any statistical test used to measure treatment effect. Nevertheless, relative difference remains invariant regardless of the number M of uninfected patients included.

The treatment efficacy for the entire sample of Boulware's study is -Rd = -(11.83-14.25) / 14.25 = 16.9%, which means that if we consider patients who presented first symptoms from days 2 to 14, then the hydroxychloroquine group decrease the average number of symptomatic patients in 16.9% when compared to placebo group in these conditions.

The treatment efficacy including only patients who had symptoms after day 5, which is the appropriate group to measure the effect of complete treatment, is -Rd = -(4.1-6.9) / 6.9 = 40.6% (see subsection 2.2, in Methods) (for complete data see Boulware et al.[1]).

**3 - Results**

In this section, we consider the same treatment of Boulware et al,[1] which included in the sample patients with a post-exposure period from 1 to 4 days before the beginning of the treatment sample, however we discriminate the sample into four sub-samples according to the number of days after exposure. The data, which have been taken from table S6 in the supplementary appendix of Boulware et al.[1], is shown in table 2 below.

Table 2: Number of patients according to time from exposure to SARS-coV-2

|  | Hydroxychloroquine group | | Placebo group | |
| --- | --- | --- | --- | --- |
| Days from exposure | Sample size | Symptomatic | Sample size | Symptomatic |
| 1 | 77 | 5 | 63 | 8 |
| 2 | 100 | 12 | 106 | 18 |
| 3 | 98 | 12 | 117 | 17 |
| 4 | 138 | 20 | 121 | 15 |

The main reason to conduct a regression study is that hypothesis tests for means, such as the Fisher's exact test adopted in Boulware et al.[1], lose vital information contained in a time series data where the effect can be measured with a greater statistical significance for the same sample size. Another important advantage, as indicated in subsection 2.2, is that regression methods are adequate to measure the effect in time heterogeneous data such as that described in table 2.

Note: The problem discussed in subsection 2.2 is not considered here, as we do not have the information necessary to proceed this specific correction. However, we do the analysis with the modifications described in subsections 2.1 and 2.3.

Let y= f(x) be the treatment effect, where the explanatory variable x=number of days from exposure to the beginning of treatment. With data from table S6, we obtain four different treatment effects, one for each x=1,2,3,4. That is, we calculate the negative relative differences of incidence of symptomatic patients of hydroxychloroquine group to placebo group as a response variable of the days from exposure to the beginning of treatment.

Hydroxychloroquine treatment efficacy and 95% confidence intervals for x= 1, 2, 3, 4 are given by 48.86% [28.97, 72.81], 29.33% [15.45, 44.15], 15.73% [-5.64, 23.06], -16.91% [-34.30, 9.53] respectively. Figure 1 shows the graph of y=f(x), for x=1,2,3,4. It also displays a simple linear regression line for this data with 95% confidence bands. Estimated slope is -21.09 with 95% confidence interval of [-32.81, -9.38]. Predicted efficacy of the treatment if applied at the same day of exposure (day 0) is 71.98% with 95% confidence interval [39.90, 100]. One-tailed p-value = 0.0081 (0.81%). Thus with 99% of confidence, we reject the hypothesis that the slope is greater than or equal to 0. This is a strong statistical evidence that hydroxychloroquine treatment reduces the proportion of symptomatic patients when used as a prophylaxis right after exposure, especially if treatment starts within 2 days.

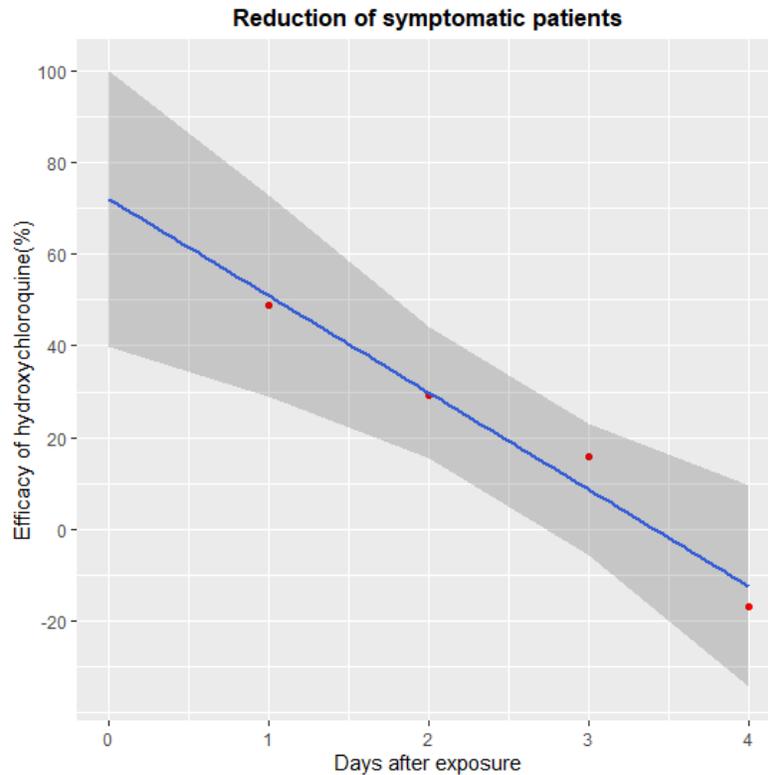

**Figure 1-** Red dots: y = hydroxychloroquine treatment efficacy and x= number of days from exposure to the beginning of treatment, for x =1,2,3,4. Blue line: Simple linear regression line for the four red dots. Dark gray shadow: 95% confidence bands for the simple linear regression line.

Next, we verify linear model suppositions. First, we assess the effect of heteroscedasticity by using weights in linear regression fit proportional to the sample sizes for x=1,2,3,4. The estimated slope is -21.63 and predicted relative difference at day 0 is 73.40%. One-tailed p-value is 0.88% in this case, which maintains the conclusions from homogeneous simple linear regression.

To assess for autocorrelation in the residuals we perform a two-tailed Durbin-Watson test using both Pan algorithm with 1000 iterations and bootstrap resampling algorithm with 10.000 replicates. Autocorrelation = -0.46, DW statistic = 2.59 and p-value = 89.8%. Thus, at 95% confidence level, the hypothesis H0 (autocorrelation = 0) is not rejected.

To assess the normality of the residuals, we obtain kurtosis = 2.09, skewness = 0.77 and perform the Shapiro-Wilkson test. The W statistic = 0.91, p-value = 46.3% and at 95% confidence level the normality hypothesis is not rejected.

# 4 - Discussion

In this study, we discussed some inaccuracies in the statistical analysis of Boulware et al.[1] We also add an original statistical analysis by adopting a different method, replacing Fisher's exact test with a simple regression analysis. There are two main reasons for this choice: first, the data of the trial is time dependent and a mean type test like Fisher's ignores this important information which eventually lead to a different conclusion; second, the number of infected asymptomatic patients, which is necessary to use Fisher's exact test, is is unknown in this data, invalidating their results.

At Boulware et al.[1], the authors analysis did not rejected the hypothesis that hydroxychloroquine effect was equal to placebo effect and they concluded that hydroxychloroquine did not prevent symptoms of Covid-19 as prophylaxis treatment. Note this conclusion cannot be made by any hypothesis test, which only states in this case there is not enough statistical evidence to refuse null hypothesis, which is different from stating the alternative hypothesis is correct. Their conclusion incorrectly states there is no evidence of efficacy, while the evidence is positive although not conclusive at 95% level with the sample size and methodology used.

Applying the modifications we have stated in sections 2 and 3, in particular using a simple linear regression method to their data, we conclude the randomized trial of Boulware et al.[1] has statistical evidence, at 99% confidence level, that hydroxychloroquine treatment is time-dependent with a negative slope. We conclude that, when applied as a prophylaxis, it can significantly reduce the relative proportion of symptomatic patients if used from 0 to 2 days after exposure to the virus (71.98% for 0 days, 48.86% for 1 day and 29.33% for 2 days). The predictive value for day 0 can be seen as lower bound for the expected hydroxychloroquine efficacy if used as a pre-exposure prophylaxis. This suggests that pre-exposure prophylaxis can be significantly effective. For 3 and 4 days, we conclude there is no statistical evidence, at 99% level, that hydroxychloroquine treatment reduces the proportion of symptomatic patients.

Moreover, our results show that the elapsed time between the exposure to the virus and the beginning of treatment is vital to the effectiveness of the antiviral use. We expect the treatment will be more effective when applied to patients in the viral replication period, before viral load reaches its peak which occurs around 5 days after symptom onset.[6] Meanwhile, if disease reaches the inflammatory period, typically after 8 days of symptoms onset and after viral load reaches its peak,[6] we can expected no or little benefit with the antiviral treatment.

Therefore, the mean time elapsed from exposure to the virus and the start of treatment in the sample may act as a lurking variable, influencing in a hidden way the efficacy of treatment. This might explain why many studies have found no statistical evidence of effectiveness of hydroxychloroquine treatment when used in hospitalized patients as most of this more severe cases had probably started treatment long after 4 days from their exposure to the virus.[3,4,6] In addition, it helps to understand why some studies have shown some positive results of hydroxychloroquine treatment as we can expect this when the proportion of patients in the beginning of the infection is higher in the sample.[3] Hence, as described by Boulware et al.,[1] two possible applications would be to apply prophylaxis to health professionals and to contacts of positive patients, since these two groups would have a greater probability to benefit from treatment.

Our results suggest there is probably little or no benefit if the treatment is used in patients infected for too long, like hospitalized severe patients. On the contrary, they also suggest infected patients may have a large benefit if treated as early as possible, mostly as pre-prophylaxis treatment where symptoms appear will have an estimated relative reduction of at least 72%.

Another important aspect is that the variable of the study, be asymptomatic or be symptomatic, is quite time sensitive. Future trials should adopt less time sensitive variables, such as the number of days each patient is symptomatic, which could measure the possible benefit of treatment for patients that have been exposed for more than 3 days before the beginning of treatment. Another common possibility is to adopt some score system to measure severity of symptoms.[1] However, score systems are difficult to be scientifically validated because they typically depend on personal judgment, whereas the variable number of days with symptoms can be more easily replicated by other studies.

Furthermore, the hydroxychloroquine prophylaxis should also be investigated with concomitant use of azithromycin and zinc[7], as also other antivirals should also be tested as prophylaxis, measuring the relative efficacy as a measure of the elapsed time after exposure to the virus to the beginning of treatment.[8]

We conducted all statistical analysis with R software, version 4.0.0.

We declare no conflict of interests.

Author: Márcio Watanabe, Ph.D
Department of Statistics, Federal Fluminense University

Address: Rua Professor Marcos Waldemar de Freitas Reis, s/n
Instituto de Matemática e Estatística – Bloco H – 3 andar – Ala A
São Domingos – 24.2010-201 – Niterói – RJ, Brasil

Email: souza_marcio@id.uff.br

# References

**1.** Boulware DR, Pullen MF, Bangdiwala AS, et al. A randomized trial of hydroxychloroquine as postexposure prophylaxis for Covid-19. N Engl J Med (2020). DOI: 10.1056/NEJMoa2016638.

**2.** Cohen MS. Editorial - Hydroxychloroquine for the Prevention of Covid-19 - Searching for Evidence. N Engl J Med (2020). DOI: 10.1056/NEJMe2020388.

**3.** Pastick KA, Okafor EC, Wang F, et al. Review: hydroxychloroquine and chloroquine for treatment of SARS-CoV-2 (COVID-19). Open Forum Infect Dis 2020; 7: ofaa130.

**4.** Rosenberg ES, Dufort EM, Udo T, et al. Association of treatment with hydroxychloroquine or azithromycin with in-hospital mortality in patients with COVID-19 in New York State. JAMA 2020 May 11 (Epub ahead of print).

**5.** Liu J, Cao R, Xu M., et al. Hydroxychloroquine, a less toxic derivative of chloroquine, is effective in inhibiting SARS-CoV-2 infection in vitro. Cell Discov 6, 16 (2020). https://doi.org/10.1038/s41421-020-0156-0

**6.** Tay, M.Z., Poh, C.M., Rénia, L.et al. The trinity of COVID-19: immunity, inflammation and intervention. Nat Rev Immunol 20,363–374 (2020). https://doi.org/10.1038/s41577-020-0311-8

**7.** Gautret, P. et al. Hydroxychloroquine and azithromycin as a treatment of COVID-19: results of an open-label non-randomized clinical trial. Int. J. Antimicrob. Agents. (2020). https://doi.org/10.1016/j.ijantimicag.2020.105949

**8.** Wang Y, Zhang D, Du G, et al. Remdesivir in adults with severe COVID-19: a randomised, double-blind, placebo-controlled, multicentre trial. Lancet 2020;395:1569-1578.